\documentclass[reprint,superscriptaddress]{revtex4-1}
\usepackage{amsmath}
\usepackage{amssymb}
\usepackage{color}
\usepackage{siunitx}
\usepackage{graphicx}
\usepackage{dcolumn}
\usepackage{bm}
\usepackage{hyperref}

\begin{document}
\title{A fully \emph{ab initio} approach to inelastic atom--surface scattering}
\author{Michelle M. Kelley}
\affiliation{Department of Physics, Cornell University, Ithaca, New York 14853, USA}
\author{Ravishankar Sundararaman}
\affiliation{Department of Materials Science \& Engineering, Rensselaer Polytechnic Institute, Troy, New York 12180, USA}
\author{Tom\'as A. Arias}
\affiliation{Department of Physics, Cornell University, Ithaca, New York 14853, USA}

\date{\today}
\begin{abstract}
We introduce a fully \textit{ab initio} theory for inelastic scattering of any atom from any surface exciting single phonons, and apply the theory to helium scattering from Nb(100). The key aspect making our approach general is a direct first-principles evaluation of the scattering atom--electron vertex. By correcting misleading results from current state-of-the-art theories, this fully \textit{ab initio} approach will be critical in guiding and interpreting experiments that adopt next-generation, non-destructive atomic beam scattering.
\end{abstract}
\maketitle
As opposed to scattering electrons or x-rays, atomic and molecular beams are non-destructive surface probes that allow for investigations of increasingly sensitive and delicate samples, pushing the scientific limits of surface-types that can be feasibly examined~\cite{Toennies1993, Benedek1994, Farias1998}. Such low-energy ($<\,$0.1~eV) beams of atoms---which do not react with or damage samples and characteristically scatter a few angstrom above surfaces---open up opportunities to study wider classes of materials including fragile biological specimens, polymers, glass, topological materials and even meta-stable or reactive surfaces that would otherwise be inaccessible~\cite{Myles2019, Holst2021,Tamtogl2021, Schmutzler2022, Auerbach2021, Myles2020,Pan2021,Chadwick2022, Lambrick2022}. Modern advances in atomic scattering techniques include the recently invented helium-atom microscopy and helium spin-echo spectroscopy, where helium is a popular choice of scatterer for reasons such as its small mass and chemical inertness~\cite{Estermann1930,Koch2008, Avidor2011, Tamtogl2015, Rotter2016, Buchner2018, Barr2016, Jones2016, Tamtogl2018}. Despite the promise of these innovative methods, there remain challenges such as low detection-efficiencies ($\sim$5--6 orders of magnitude less than EELS~\cite{Benedek2018}), which can be remedied by integrating over many beam pulses but requires re-cleaning and maintaining the surface throughout the integration process. 

Theoretical predictions of atomic scattering signatures are critical. Such predictions are not only necessary to guide the experimental measurement process with its low detection-efficiency, but also to interpret the resulting data. As we will show below, existing semi-empirical theories are often misrepresentative---downplaying or completely missing distinctive features while overemphasizing others---which makes identifying the fundamental underlying processes extremely challenging. Unfortunately, no fully \textit{ab initio} method, which computes scattering directly from first principles, has yet been available to guide and interpret atom--surface scattering experiments.

Advances in the theory behind atom--surface scattering have mostly centered around developing different model potentials for the distorted-wave Born approximation~\cite{Benedek2018,Hubbard1983,Eichenauer1987,Stiles1988,Gumhalter2001, Siber2001}. One particularly important development came after the first observation of the anomalous phonon resonance~\cite{Doak1983}, which is now understood as a feature common to metallic surfaces~\cite{Benedek2018}. The interpretation of this surface-phonon resonance established that helium atoms scatter off of the surface free-electron density as opposed to individual surface atoms, meaning inelastic atom--surface scattering contains information on how electron--phonon interactions manifest at surfaces~\cite{Chis2008,Benedek2010}. A theory for inelastic helium-atom scattering (HAS) incorporating the underlying electron--phonon interactions was finally formulated in cutting-edge work from 2011~\cite{Sklyadneva2011}, which proposed that inelastic HAS probabilities are approximately proportional to electron--phonon coupling (EPC) strengths $\lambda_{\mathbf q \nu}$ and led to an important sequence of papers~\cite{Tamtogl2013,Benedek2014,Manson2016,Benedek2022,Manson2022a}. Experimentally accessing these EPC strengths is important because these fundamental parameters quantify most properties of conventional superconductors~\cite{Schrieffer1957,Grimvall1983}, including $T_\textrm c$~\cite{McMillan1968,Allen1975}. However, the idea that inelastic HAS probabilities are proportional to $\lambda_{\mathbf q \nu}$ is an oversimplification. That proportionality would imply that the underlying helium--electron interactions can be trivially factored out when computing inelastic HAS probabilities, but this is not the case. Moreover, experiments show different scattering behaviors depending on the choice of probe particle~\cite{Petersen1996, Minniti2012}, a result which demands a universal theory capable of discerning subtle differences among distinct types of scattering species. A complete understanding of the physics encompassed in atom--surface scattering ultimately requires a fully \textit{ab initio} framework to calculate explicit interactions between the probe atom and surface electrons. These interactions comprise a fundamental component in atom--surface scattering that have been mostly ignored until now and never before computed directly from first principles. 
\begin{figure}[t!]
\includegraphics[width=\columnwidth]{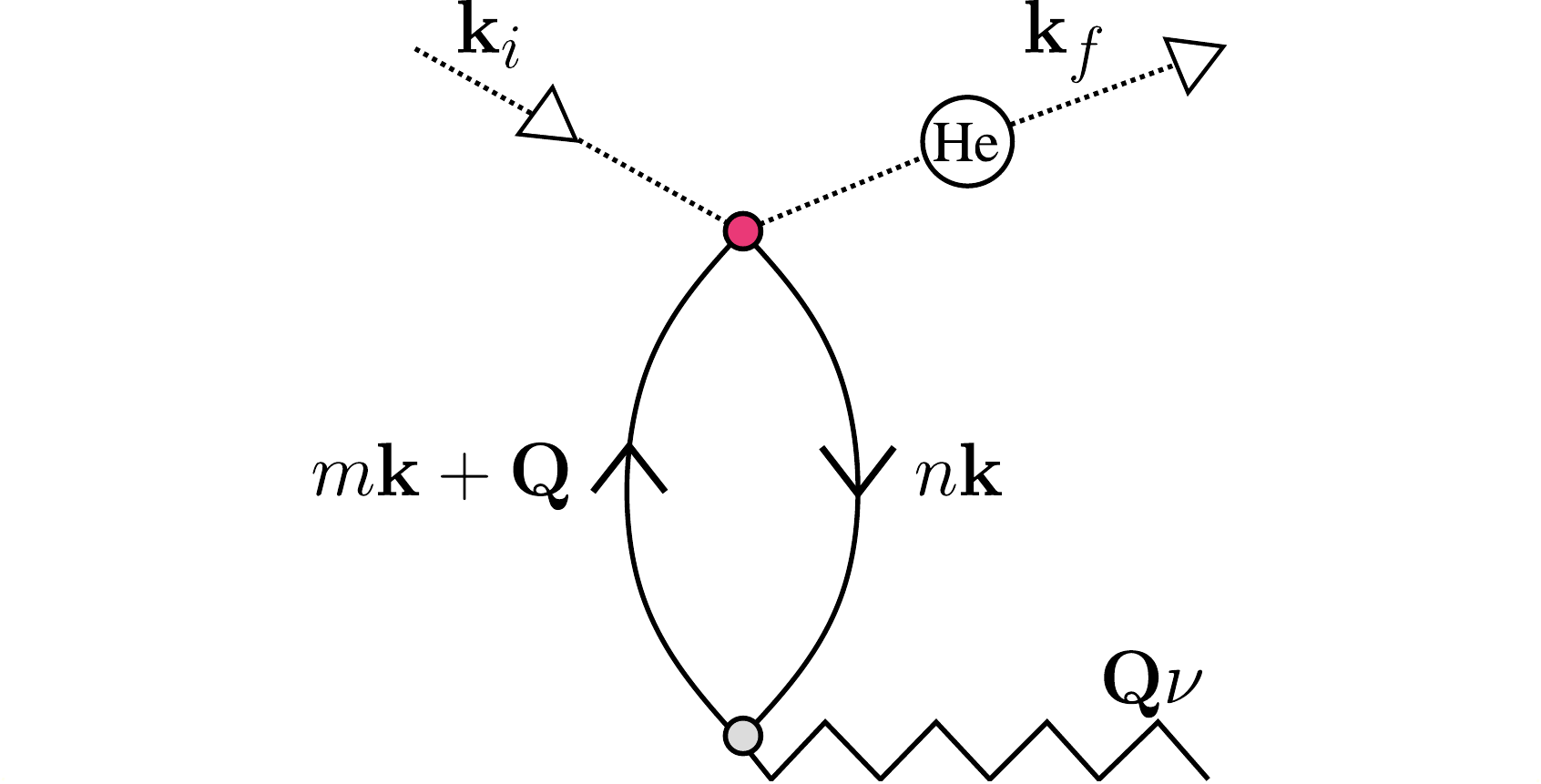}
\caption{A helium atom (dotted lines) with initial and final momenta $\mathbf k_{i,f}$ indirectly excites a surface phonon (jagged line) mode $\nu$  with lateral momentum $\mathbf Q$ via an electron--hole pair excitation (solid lines) from bands $n$ and $m$ with momenta $\mathbf k$ and $\mathbf{k+Q}$. Intersecting lines show the helium--electron vertex (dark magenta) and the electron--phonon vertex (light gray).}
\label{fig:diagram}
\end{figure}

Here, we introduce an entirely \textit{ab initio} framework for inelastic atom--surface scattering. This work provides a new approach to predict HAS intensities and reports the first \textit{ab initio} evaluation of the helium atom--electron vertex (see Fig.~\ref{fig:diagram}). Moreover, this vertex is fundamental to all atom--surface scattering processes---well beyond the leading-order inelastic process in Fig.~\ref{fig:diagram}---including elastic scattering and other inelastic events like acoustic surface-plasmon excitations~\cite{Benedek2021}. We apply our method computing Fig.~\ref{fig:diagram} from first principles to Nb(100) and compare to previously published HAS measurements to demonstrate the validity of our new approach~\cite{Hulpke1992}. Additionally, we demonstrate the superiority of our approach over two lower levels of theory. The first level corresponds to the most commonly used simplification of the distorted-wave Born approximation. The second level gives the current state-of-the-art model relating HAS probabilities to EPC strengths~\cite{Sklyadneva2011, Tamtogl2013, Benedek2014, Manson2016, Benedek2022, Manson2022a}. For a more generous comparison, we slightly amend the framework of the latter method to avoid approximations to EPC and compute these interactions explicitly from first principles instead~\cite{McMillan2022}. While this work focuses on helium, this approach is easily applied to any species of scattering atom or molecule.

\emph{Theoretical framework.---}From quantum scattering theory, the helium--electron vertex from Fig.~\ref{fig:diagram} corresponding to a helium atom at $\mathbf r'$ can be written as
\begin{equation}
h_{m\mathbf{k+Q},n\mathbf k}(\mathbf r')=\int d\mathbf r'\psi^\dagger_{n\mathbf k}(\mathbf r)\Delta V_\textrm{He}(\mathbf r,\mathbf r')\psi_{m\mathbf{k+Q}}(\mathbf r),
\label{eq:h_r}
\end{equation}
where $\Delta V_\textrm{He}(\mathbf r,\mathbf r')$ gives the perturbing potential from the addition of a helium atom at $\mathbf r'$, and $\mathbf r$ denotes the electronic coordinate. Here, we adopt a common convention to specify lateral coordinates using capital letters, i.e. $\mathbf R \equiv  r_x \hat{x}+r_y\hat{y}$ and $\mathbf Q \equiv  q_x \hat{x}+q_y\hat{y}$.  

\begin{figure}[t]
\includegraphics[width=\columnwidth]{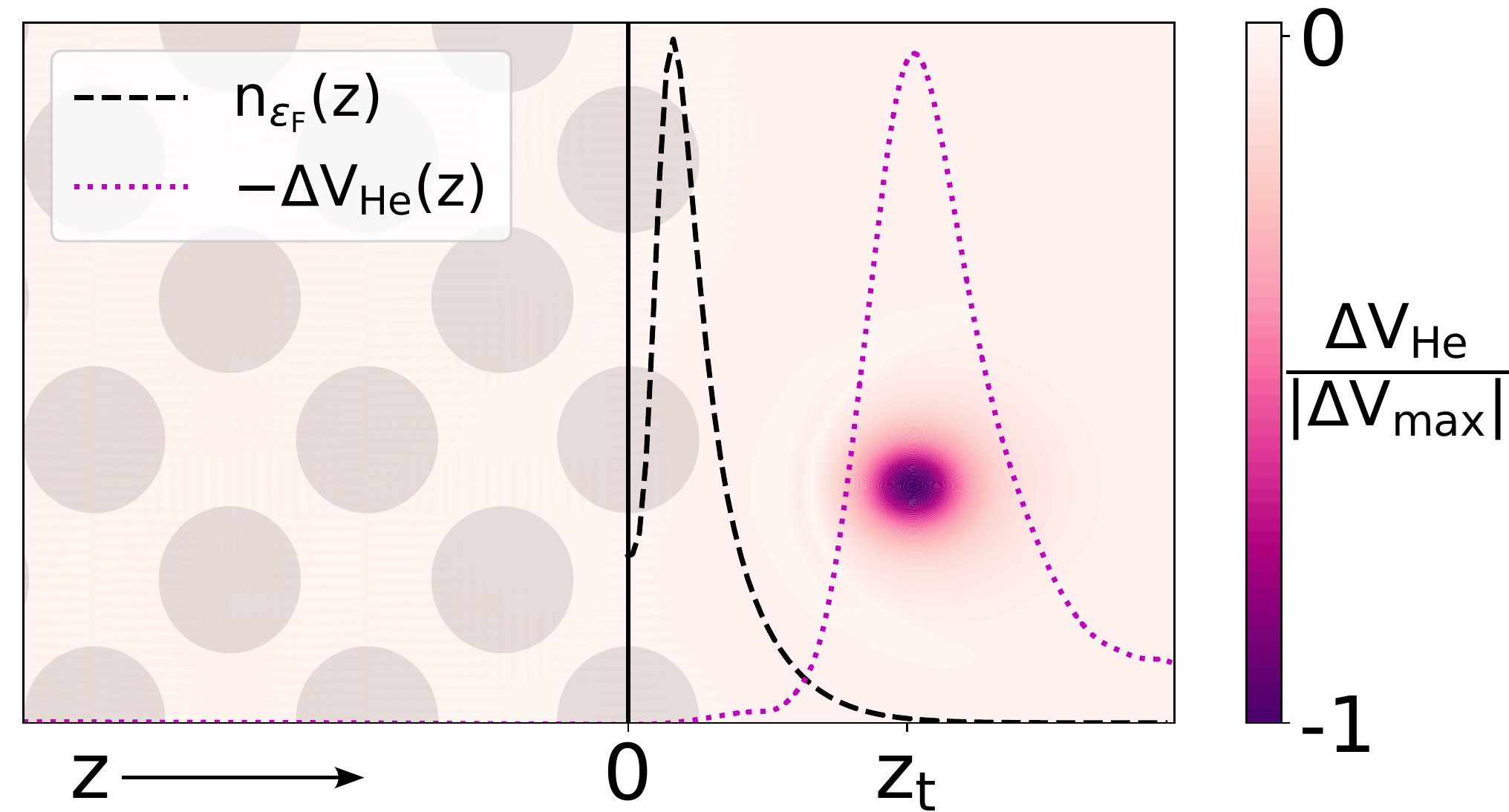}
\caption{Contour color-map showing the perturbing potential in the plane of a helium atom at its estimated turning point $z_\textrm t$ from Nb(100). The magenta dotted-line shows the helium potential, averaged over planes, as a function of distance from the surface. Gray circles show the $z$-locations of niobium atoms, with the top plane of atoms defining $z=0$. The black dashed-line shows the planar-averaged density of Fermi-level electrons at an electronic temperature of 900 K.}
\label{fig:dV}
\end{figure}
Figure~\ref{fig:dV} shows a contour plot of $\Delta V_\textrm{He}(\mathbf r,\mathbf r')$ for a helium atom at its estimated turning point from the Nb(100) surface ($z_t\approx3.4$~\AA~for $E_{iz}\approx 18$~meV, comparable to measurements from \cite{Hulpke1992}). To gain insight on the extent of the helium--electron interaction, Fig.~\ref{fig:dV} also shows the perturbing potential and density of Fermi-level electrons averaged over planes. Using Eq.~\eqref{eq:h_r}, the HAS matrix element corresponding to the scattering diagram in Fig.~\ref{fig:diagram} for a helium atom at $\mathbf r'$ becomes
\begin{equation}
\begin{split}
M_{\mathbf Q \nu}^\textrm{abs/em}(\mathbf r')=\sum_{n,m}\int &\frac{d\mathbf k}{(2\pi)^3}g_{n\mathbf k,m\mathbf{k+Q}}^{\mathbf Q\nu}h_{m\mathbf{k+Q} ,n\mathbf{k}}(\mathbf r')\\ \times &\frac{f_{n\mathbf k}-f_{n\mathbf{k+Q}}}{\epsilon_{n\mathbf k}-\epsilon_{n\mathbf{k+Q}}\pm (\omega_{\mathbf Q\nu}+i\eta)},
\end{split}
\label{eq:M_r}
\end{equation}
where $g_{n\mathbf k,m\mathbf{k+Q}}^{\mathbf Q\nu}$ gives the electron--phonon vertex, $f_{n\mathbf k}$ indicates a Fermi distribution for an electronic state with energy $\epsilon_{n \mathbf k}$, and the plus or minus sign in the denominator of the matrix element is for phonon absorption or emission, respectively. The expression in Eq.~\eqref{eq:M_r} is analogous to the familiar expression for the phonon linewidth but with two distinctions: one electron--phonon vertex is replaced with a helium--electron vertex, and we now must consider the full complex expression rather than just the imaginary component of a self-energy diagram.

\begin{figure*}[t!]
\centering
\includegraphics[width=0.96\textwidth]{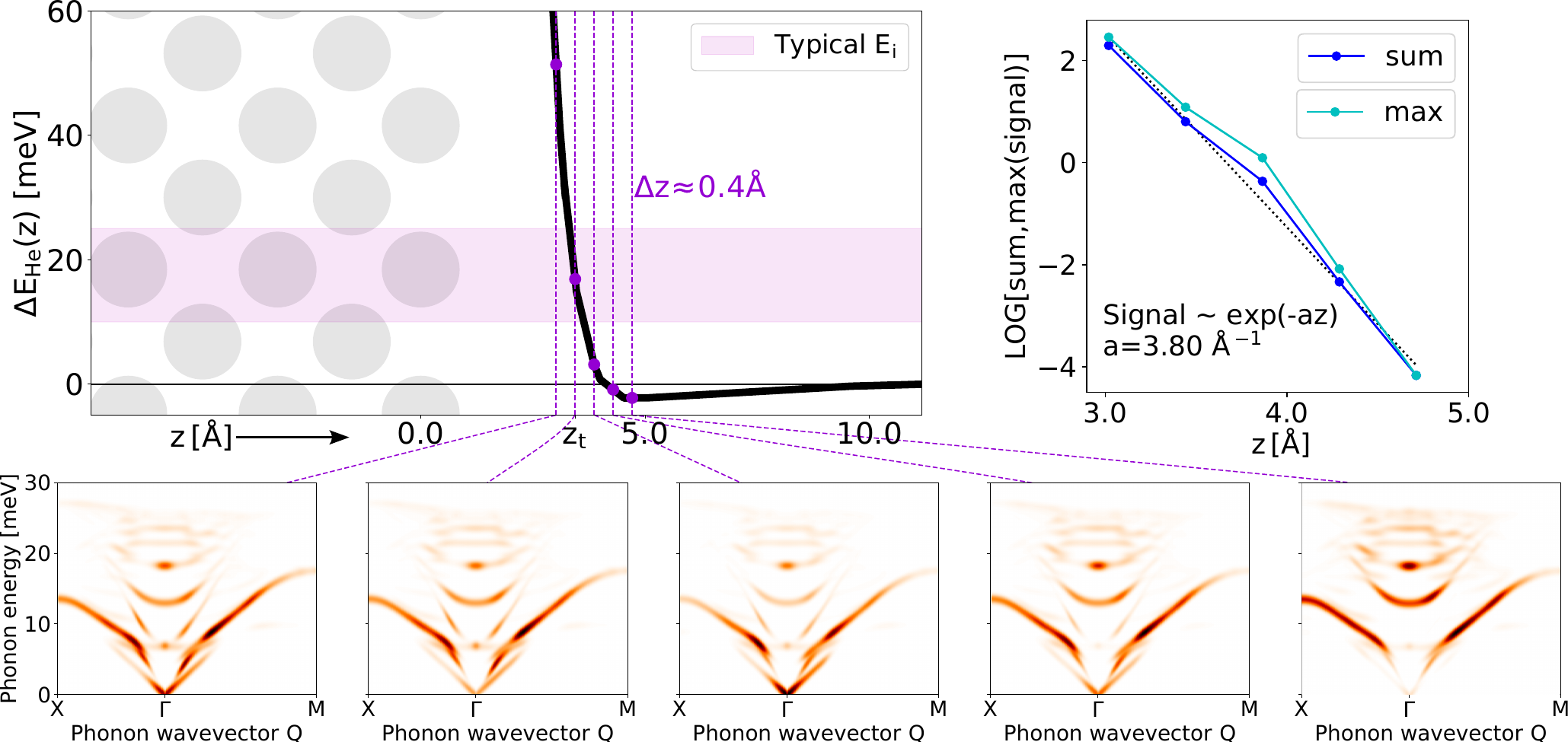}
\caption{Top-left panel shows the interaction energy profile $\Delta E_\textrm{He}$ between a helium atom and the Nb(100) surface (black curve), the shaded region depicts typical energies for incoming helium atoms, and gray circles show the $z$-locations of niobium atoms, with the top plane of atoms defining $z=0$. Bottom panel shows the predicted HAS signal intensities $|M_{\mathbf Q \nu}(z')|^2$ corresponding to helium atoms at $z$-locations given by the violet vertical-lines, spaced uniformly at approximately 0.4~\AA~apart. The top-right panel shows the natural log of the total and maximum of each signal from the bottom panel as a function of distance from the Nb(100) surface and includes the result of a line of best fit to extract the exponential decay of the signal.}
\label{fig:decay}
\end{figure*}

\begin{figure}[t]
\centering
\includegraphics[width=0.96\columnwidth]{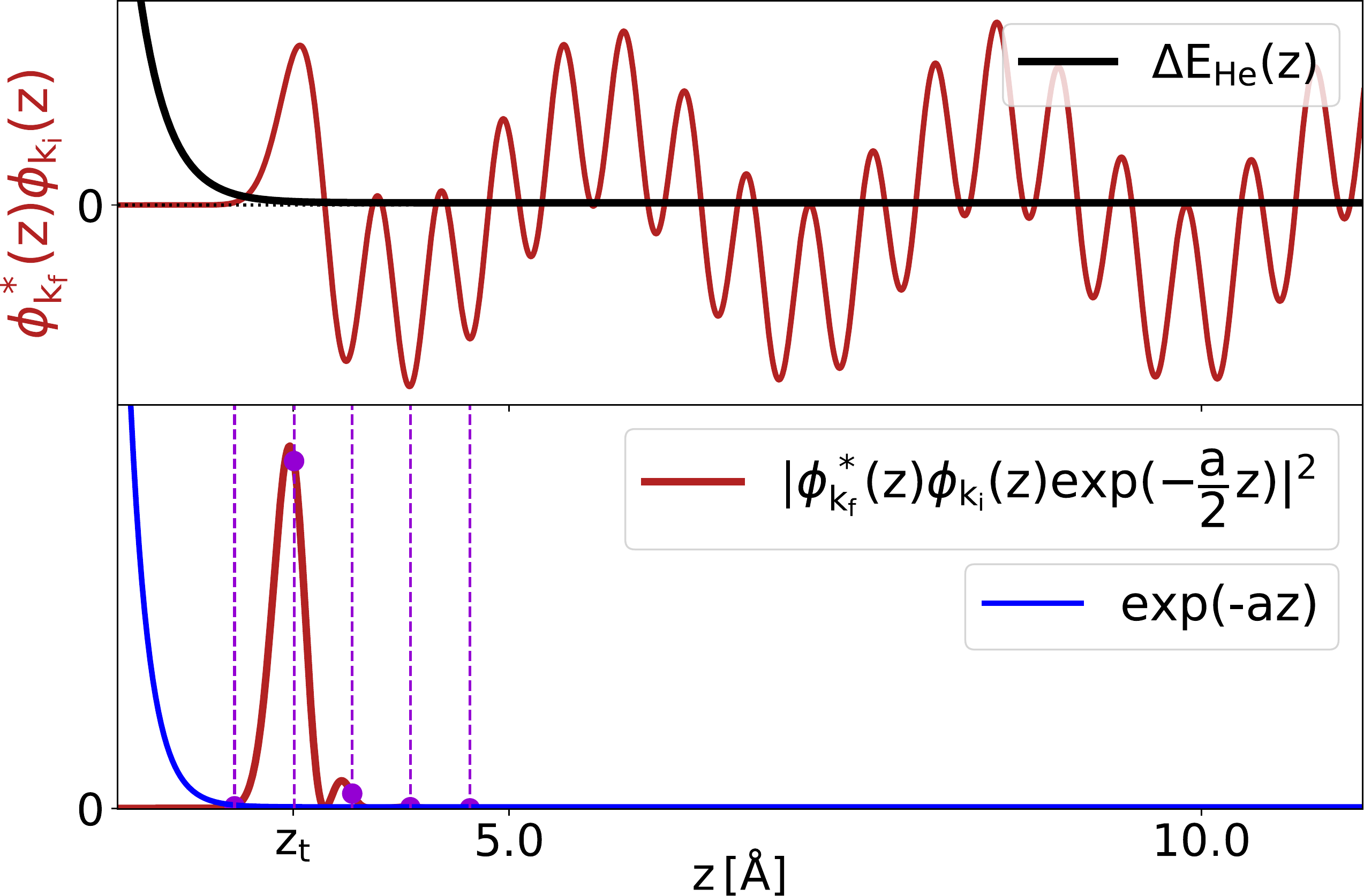}
\caption{Top panel illustrates incoming and outgoing helium atom $z$-wavefunctions (red curves) from potential $V(z)=\Delta E_\textrm{He}(z)$. Bottom panel combines the wavefunctions with the exponential decay of the helium--surface interaction (blue curve). The violet vertical-lines in the bottom panel give the same $z$-coordinates of the helium atoms shown in Fig.~\ref{fig:decay}.}
\label{fig:wfs}
\end{figure}

\begin{figure}[t]
\centering
\includegraphics[width=\columnwidth]{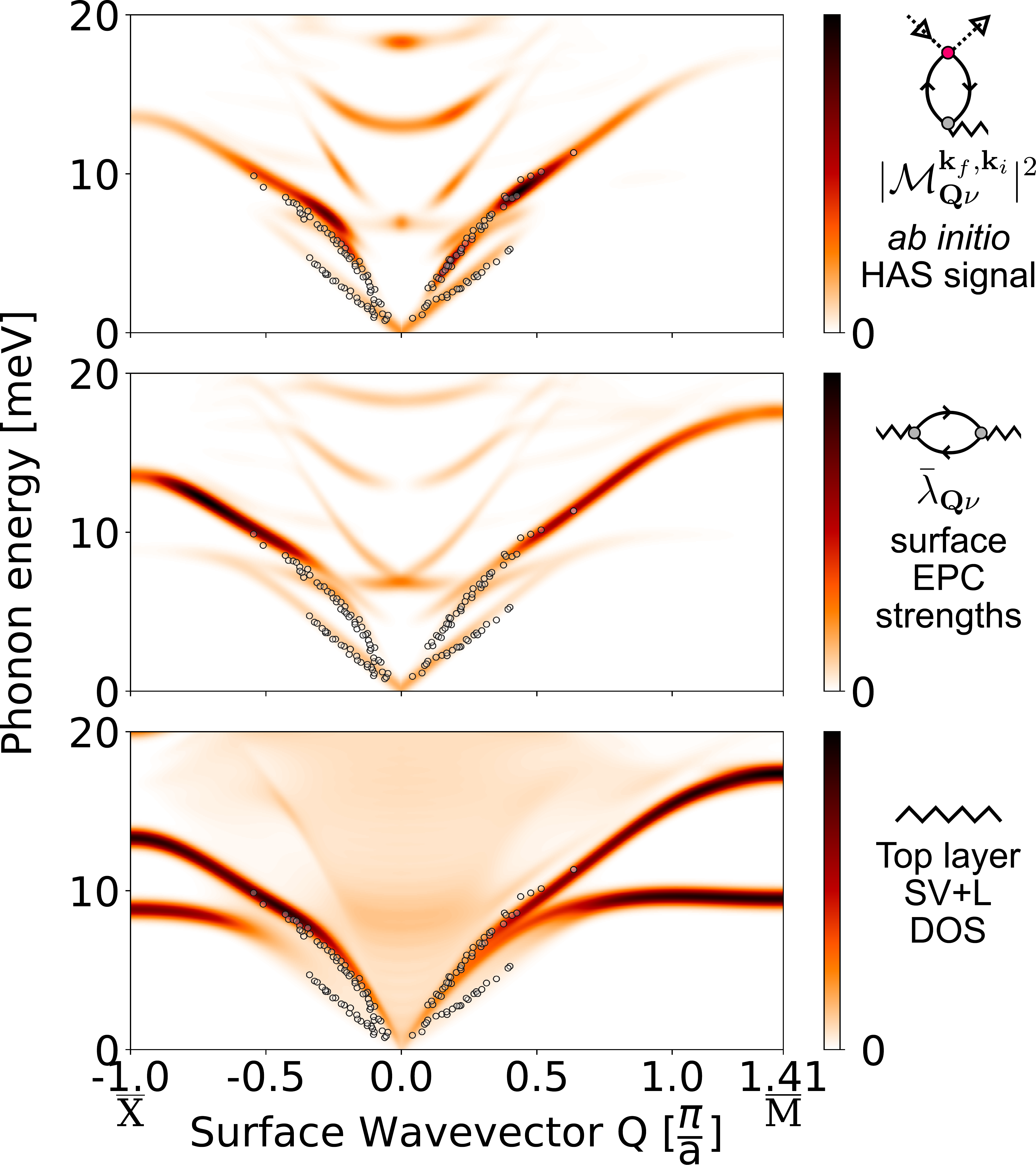}
\caption{Color maps showing predicted inelastic helium atom scattering (HAS) signals from Nb(100) at three levels of theory. Color-bar labels show the Feynman diagram elements considered at each level. The bottom panel considers only phonons to show the top-layer phonon density-of-states for shear vertical (SV) and longitudinal (L) polarizations. The middle panel incorporates the electron--phonon vertex to give the inelastic HAS predictions estimated from surface electron-phonon coupling (EPC) strengths~\cite{McMillan2022}. The top panel includes both the electron--phonon and helium--electron vertices, giving the final prediction of the fully \textit{ab initio} HAS calculation from Eq.~\eqref{eq:finalHAS}. Black circles in each panel correspond to measured helium-scattering data~\cite{Hulpke1992}.}
\label{fig:HAS}
\end{figure}
The total scattering matrix element integrates Eq.~\eqref{eq:M_r} over the helium coordinate $\mathbf r'$, weighted by the helium atom's wavefunction
\begin{equation}
\begin{split}
\mathcal M_{\mathbf Q \nu,\textrm{abs/em}}^{\mathbf k_i,\mathbf k_f}=\sum_\mathbf{G}\delta&(\mathbf{Q+G}-\Delta \mathbf K_{fi}) \\ \times &\int_\Omega d\mathbf r' \Phi^{*\mathbf k_f}_{\textrm{He}}(\mathbf r') \Phi^{\mathbf k_i}_{\textrm{He}}(\mathbf r') M_{\mathbf Q \nu}^\textrm{abs/em}(\mathbf r').%
\end{split}
\label{eq:Mfi}
\end{equation}
The total inelastic scattering probability for a helium atom ($\mathbf k_i\to\mathbf k_f$) that absorbs or emits one phonon is ultimately calculated from
\begin{equation}
\begin{split} 
     \textrm{Absorption:~~} &|\mathcal M_{\mathbf Q \nu,\textrm{abs}}^{\mathbf k_i,\mathbf k_f}|^2\times n(\omega_{\mathbf Q \nu})  \\
     \textrm{Emission:~~} &|\mathcal M_{\mathbf Q \nu,\textrm{em}}^{\mathbf k_i,\mathbf k_f}|^2\times \left(n\left(\omega_{\mathbf Q \nu}\right)+1\right),
   \end{split}
   \label{eq:abs_em}
\end{equation}
where $n(\omega_{\mathbf Q\nu})$ gives the boson occupancy for the phonon mode involved in the collision. Equations~{\eqref{eq:h_r}--\eqref{eq:abs_em}} provide our general framework, and we will now demonstrate how the approach simplifies for low surface corrugation.


\emph{Surfaces with low corrugation.---}For surfaces with low corrugation like Nb(100), the lateral coordinate in Eq.~\eqref{eq:Mfi} factors out through $M_{\mathbf Q \nu}(\mathbf r')=M_{\mathbf Q \nu}(z')$, and the helium atom wavefunction can be approximated with $\Phi^{\mathbf k}_{\textrm{He}}(\mathbf r')\approx e^{i \mathbf K \cdot \mathbf R'}\phi_{\mathbf k}(z')$.
Now, the integral over the lateral coordinate simplifies conveniently, reducing to sinc functions for systems of orthorhombic symmetry. As a result, the last remaining piece to evaluate is an integral of $M_{\mathbf Q \nu}(z')$ over the helium atom's $z$-coordinate.

The helium atom's $z$-coordinate is the central variable determining the atom's interaction with the Nb(100) surface. The black curve in the top-left panel of Fig.~\ref{fig:decay} shows the interaction energy profile of a helium atom as a function of distance from the surface. We investigate the position-dependent HAS matrix element $M_{\mathbf Q \nu}(z')$ by sampling a helium atom at five distances spaced uniformly from the surface (Fig.~\ref{fig:decay}; violet vertical-lines) and compute the corresponding HAS signal intensity $|M_{\mathbf Q \nu}(z')|^2$ for each case (Fig.~\ref{fig:decay}; bottom panel). Because helium scatters from the surface electron-density, the HAS matrix element weakens as the helium atom recedes the surface. Indeed, we find a well-defined exponential decay constant of $3.8$~\AA$^{-1}$ for the position-dependent HAS signal (Fig.~\ref{fig:decay}; top-right panel). The interaction clearly strengthens as helium approaches the surface, but the total scattering matrix element requires the atomic wavefunction to complete the integration in Eq.~\eqref{eq:Mfi}.
The helium atom's $z$-wavefunction can be approximated by solving a one-dimensional Schr\"{o}dinger equation with a potential imposed by the helium atom's interaction energy with the surface $V(z')=\Delta E_\textrm{He}(z')$. Figure~\ref{fig:wfs} shows example incoming and outgoing $z$-wavefunctions from this potential and the integrand of the HAS matrix element from Eq.~\eqref{eq:Mfi} as a function of distance from the surface. These results reveal that the HAS signal is dominated by the contribution of the helium atom at its turning point $z_\textrm t$. We find this to be the case for all trial wavefunctions that we have considered, indicating that relative HAS intensities can be well-estimated from 
\begin{equation}
\begin{split}
\mathcal M_{\mathbf Q \nu}^{\mathbf k_i,\mathbf k_f}\approx&\sum_\mathbf{G}\delta(\mathbf{Q+G}-\Delta \mathbf K_{fi})\\ &\times M_{\mathbf Q \nu}(z_\textrm t)\, \textrm{sinc}\left(\frac{\Delta K_x R_x}{2} \right)\textrm{sinc}\left(\frac{\Delta K_y R_y}{2} \right),
\end{split}
\label{eq:finalHAS}
\end{equation}
where $\textrm{sinc}(x)\equiv \frac{\sin(x)}{x}$ and $R_{i}$ are surface lattice vectors.

\emph{Computational methods.---}To study Nb(100), we perform density-functional theory (DFT) calculations using open-source planewave software JDFTx~\cite{Payne1992, Sundararaman2017}. We apply norm-conserving pseudopotentials~\cite{Schlipf2015} and calculate the electronic states for the outer electrons of niobium (${4p^65s^24d^3}$) and helium ($1s^2$) at an effective temperature of 20~milli-Hartree using a Fermi function to determine electronic occupancies. To approximate the exchange-correlation energy~\footnote{A brief discussion regarding the choice of xc-functional approximation is given in the Supplemental Material, which includes Refs.~\cite{Giustino2017, Marzari2012, Vega2018}.}, we apply the Perdew–Burke–Ernzerhof functional revised for solids (PBEsol)~\cite{Perdew2008}. All calculations employ planewave cutoff energies of 30~Hartree and 200~Hartree for the electronic wavefunctions and density, respectively. We calculate a 10-layer slab of niobium with (100) surface termination in a cell that is 42.33~\AA~long along the surface-normal direction and truncate Coulomb potentials to increase the accuracy of calculated surface properties~\cite{Sundararaman2013}. We calculate a lateral lattice constant for Nb(100) at 3.30~\AA, in good agreement with the experimental measurement of 3.29~\AA~\cite{Veit2019}. Interatomic force constant matrices and helium interactions are calculated in a 3$\times$3$\times$1 supercell with a $\mathbf k$-space sampling density equivalent to the unit cell's sampling of 12$\times$12$\times$1 $\mathbf k$-points. Finally, we transform into a maximally-localized Wannier function basis to interpolate helium--electron and electron--phonon scattering processes at arbitrary $\mathbf k$ and densely sample the Brillouin zone to accurately evaluate scattering integrals~\cite{Marzari1997,Giustino2007, Brown2016, Brown2016a, Xu2021,Xu2021a}.

%
\emph{Results and discussion.---}Figure~\ref{fig:HAS} shows predictions for inelastic HAS intensities at three levels of theory and compares the predictions to inelastic HAS measurements for Nb(100)~\cite{Hulpke1992}. The bottom panel depicts the least refined estimate that merely looks at the top-layer phonon density-of-states (DOS), after inserting 60 bulk dynamical matrix layers into the 10-layer Nb(100) slab, for shear vertical (SV) and longitudinal (L) polarizations. These two polarizations are the ones most commonly measured in HAS experiments and included in the distorted-wave Born approximation~\cite{Benedek2018}. The middle panel of Fig.~\ref{fig:HAS} illustrates the current state-of-the-art model estimating inelastic HAS probabilities to be proportional to surface EPC strengths $\lambda_{\mathbf Q \nu}$~\cite{Sklyadneva2011}, but we refine this model to improve predictions by calculating electron--phonon matrix elements \textit{ab initio}~\cite{McMillan2022}. The top panel of Fig.~\ref{fig:HAS} gives the highest level of theory, corresponding to the expression from Eq.~\eqref{eq:finalHAS}, which now considers the full scattering diagram from Fig.~\ref{fig:diagram} and incorporates both the electron--phonon and helium--electron vertices \textit{ab initio}.

Before assessing the predictions given at each level of theory, it is necessary to first understand HAS measurements to interpret the data. The density of measured points reflects the detectability of phonon modes, influenced by the intrinsic availability of the modes and experimental conditions. Atoms are big and slow relative to electrons, and inelastic scattering signals will be ``cut-off'' beyond certain values of phonon energy and wavevector because the atom is unable to excite those modes~\cite{hulpke1992helium}. This cut-off factor is not absolute and depends on kinematic factors of the scattering atom that will affect the resulting signal-to-noise ratio~\cite{Meyer1982,Schinke1983}. Inelastic intensities are strongest near $\Gamma$, and data collection proceeds along the observed phonon branches until the signal becomes undetectable~\cite{Celli1984}. Hence, data-points abruptly stopping along a branch indicates the locations where the signal became undetectable.

As expected, the top-layer phonon DOS shown in the bottom panel in Fig.~\ref{fig:HAS} provides the crudest estimate to inelastic HAS signals. This prediction strikingly misses the lower measured mode, incorrectly predicts signals to be strongest towards the edges of the Brillouin zone, and overall illustrates why examining merely the phonon DOS conveys an inadequate picture of inelastic HAS signals. Next, the prediction given by the surface EPC strengths successfully captures both measured surface-phonon modes, but the signal predicted for both modes continues after the data stop, there appears to be spurious signal between the two measured modes, and the signal for the upper mode notably increases after most of the data stop and is strongest where no data have been measured. Finally, all of these incorrect features from the above approaches are corrected in the top panel of Fig.~\ref{fig:HAS} showing the fully \textit{ab initio} HAS analysis. Upon properly including the helium--electron interaction, the predicted signal for the lower mode decays in remarkable agreement with the measured data, there is hardly any extra signal predicted between the two measured modes, and even though there is still some signal predicted after the data stop in the upper branch, the signal nonetheless decays after the data stop and the most intense regions align well with measurements.


The analysis above demonstrates the critical importance of a first principles evaluation of the helium atom--electron vertex in predicting and understanding the inelastic helium-atom scattering process. This work provides a general framework for computing inelastic atom--surface scattering and produces results of high accuracy. This theoretical approach will provide the needed guidance for the performance and interpretation of next-generation experiments using atomic beam scattering as a non-destructive probe of sensitive surfaces.



We thank Caleb Thompson, Michael Van Duinen, and Steven Sibener for useful discussions regarding helium-scattering experiments. This work was supported by the US National Science Foundation under award PHY-1549132, the Center for Bright Beams.

\bibliographystyle{myunsrt}
\bibliography{HAS}

\begin{thebibliography}{10}

\bibitem{Toennies1993}
J.~P. Toennies,
\newblock {\em J. Phys. Condens. Matter} \href{http://dx.doi.org/10.1088/0953-8984/5/33A/003}{{\bf 5} 24--40} (1993).

\bibitem{Benedek1994}
G.~Benedek and J.~P. Toennies,
\newblock {\em Surf. Sci.} \href{http://dx.doi.org/10.1016/0039-6028(94)90683-1}{{\bf 299-300} 587--611} (1994).

\bibitem{Farias1998}
D.~Far{\'{i}}as and K.~H. Rieder,
\newblock {\em Reports Prog. Phys.} \href{http://dx.doi.org/10.1088/0034-4885/61/12/001}{{\bf 61} 1575--1664} (1998).

\bibitem{Myles2019}
T.~A. Myles, S.~D. Eder, M.~G. Barr, A.~Fahy, J.~Martens  and P.~C. Dastoor,
\newblock {\em Sci. Rep.} \href{http://dx.doi.org/10.1038/s41598-018-36373-5}{{\bf 9} 1--10} (2019).

\bibitem{Holst2021}
B.~Holst, G.~Alexandrowicz, N.~Avidor, G.~Benedek, G.~Bracco, W.~E. Ernst, D.~Far{\'{i}}as, A.~P. Jardine, K.~Lefmann, J.~R. Manson, R.~Marquardt, S.~M. Art{\'{e}}s, S.~J. Sibener, J.~W. Wells, A.~Tamt{\"{o}}gl  and W.~Allison,
\newblock {\em Phys. Chem. Chem. Phys.} \href{http://dx.doi.org/10.1039/d0cp05833e}{{\bf 23} 7653--7672} (2021).

\bibitem{Tamtogl2021}
A.~Tamt{\"{o}}gl, A.~Ruckhofer, D.~Campi, W.~Allison  and W.~E. Ernst,
\newblock {\em Phys. Chem. Chem. Phys.} \href{http://dx.doi.org/10.1039/d0cp05388k}{{\bf 23} 7637--7652} (2021).

\bibitem{Schmutzler2022}
S.~J. Schmutzler, A.~Ruckhofer, W.~E. Ernst  and A.~Tamt{\"{o}}gl,
\newblock {\em Phys. Chem. Chem. Phys.} \href{http://dx.doi.org/10.1039/d1cp05284e}{{\bf 24} 9146--9155} (2022).

\bibitem{Auerbach2021}
D.~J. Auerbach, J.~C. Tully  and A.~M. Wodtke,
\newblock {\em Nat. Sci.} \href{http://dx.doi.org/10.1002/ntls.10005}{{\bf 1} 1--42} (2021).

\bibitem{Myles2020}
T.~A. Myles, A.~Fahy, J.~Martens, P.~C. Dastoor  and M.~G. Barr,
\newblock {\em Meas. J. Int. Meas. Confed.} \href{http://dx.doi.org/10.1016/j.measurement.2019.107263}{{\bf 151} 107263} (2020).

\bibitem{Pan2021}
P.~Pan, M.~Debiossac  and P.~Roncin,
\newblock {\em Phys. Rev. B} \href{http://dx.doi.org/10.1103/PhysRevB.104.165415}{{\bf 104} 165415} (2021).

\bibitem{Chadwick2022}
H.~Chadwick and G.~Alexandrowicz,
\newblock {\em Phys. Chem. Chem. Phys.} \href{http://dx.doi.org/10.1039/d2cp01372j}{{\bf 24} 14198--14208} (2022).

\bibitem{Lambrick2022}
S.~M. Lambrick, M.~Bergin, D.~J. Ward, M.~Barr, A.~Fahy, T.~Myles, A.~Radi{\'{c}}, P.~C. Dastoor, J.~Ellis  and A.~P. Jardine,
\newblock {\em Phys. Chem. Chem. Phys.} \href{http://dx.doi.org/10.1039/d2cp01951e}{{\bf 61} 26539--26546} (2022).

\bibitem{Estermann1930}
I.~Estermann and O.~Stern,
\newblock {\em Zeitschrift f{\"{u}}r Phys.} \href{http://dx.doi.org/10.1007/BF01340293}{{\bf 61} 95--125} (1930).

\bibitem{Koch2008}
M.~Koch, S.~Rehbein, G.~Schmahl, T.~Reisinger, G.~Bracco, W.~E. Ernst  and B.~Holst,
\newblock {\em J. Microsc.} \href{http://dx.doi.org/10.1111/j.1365-2818.2007.01874.x}{{\bf 229} 1--5} (2008).

\bibitem{Avidor2011}
N.~Avidor, H.~Hedgeland, G.~Held, A.~P. Jardine, W.~Allison, J.~Ellis, T.~Kravchuk  and G.~Alexandrowicz,
\newblock {\em J. Phys. Chem. A} \href{http://dx.doi.org/10.1021/jp200221b}{{\bf 115} 7205--7209} (2011).

\bibitem{Tamtogl2015}
A.~Tamt{\"{o}}gl, E.~Bahn, J.~Zhu, P.~Fouquet, J.~Ellis  and W.~Allison,
\newblock {\em J. Phys. Chem. C} \href{http://dx.doi.org/10.1021/acs.jpcc.5b08284}{{\bf 119} 25983--25990} (2015).

\bibitem{Rotter2016}
P.~Rotter, B.~A. Lechner, A.~Morherr, D.~M. Chisnall, D.~J. Ward, A.~P. Jardine, J.~Ellis, W.~Allison, B.~Eckhardt  and G.~Witte,
\newblock {\em Nat. Mater.} \href{http://dx.doi.org/10.1038/nmat4575}{{\bf 15} 397--400} (2016).

\bibitem{Buchner2018}
C.~B{\"{u}}chner, S.~D. Eder, T.~Nesse, D.~Kuhness, P.~Schlexer, G.~Pacchioni, J.~R. Manson, M.~Heyde, B.~Holst  and H.~J. Freund,
\newblock {\em Phys. Rev. Lett.} \href{http://dx.doi.org/10.1103/PhysRevLett.120.226101}{{\bf 120} 226101} (2018).

\bibitem{Barr2016}
M.~Barr, A.~Fahy, J.~Martens, A.~P. Jardine, D.~J. Ward, J.~Ellis, W.~Allison  and P.~C. Dastoor,
\newblock {\em Nat. Commun.} \href{http://dx.doi.org/10.1038/ncomms10189}{{\bf 7} 1--5} (2016).

\bibitem{Jones2016}
A.~Jones, A.~Tamt{\"{o}}gl, I.~Calvo-Almaz{\'{a}}n  and A.~Hansen,
\newblock {\em Sci. Rep.} \href{http://dx.doi.org/10.1038/srep27776}{{\bf 6} 1--11} (2016).

\bibitem{Tamtogl2018}
A.~Tamt{\"{o}}gl, M.~Pusterhofer, M.~Bremholm, E.~M. Hedegaard, B.~B. Iversen, P.~Hofmann, J.~Ellis, W.~Allison, S.~Miret-Art{\'{e}}s  and W.~E. Ernst,
\newblock {\em Surf. Sci.} \href{http://dx.doi.org/10.1016/j.susc.2018.02.006}{{\bf 678} 25--31} (2018).

\bibitem{Benedek2018}
G.~Benedek and J.~P. Toennies,
\newblock {\em {Atomic Scale Dynamics at Surfaces}} volume~63,
\newblock (2018).

\bibitem{Hubbard1983}
L.~M. Hubbard, S.~Shi  and W.~H. Miller,
\newblock {\em The Journal of Chemical Physics} {\bf 78} 2381--2387 (1983).

\bibitem{Eichenauer1987}
D.~Eichenauer, U.~Harten, J.~P. Toennies  and V.~Celli,
\newblock {\em The Journal of Chemical Physics} {\bf 86} 3693--3710 (1987).

\bibitem{Stiles1988}
M.~D. Stiles and J.~W. Wilkins,
\newblock {\em Phys. Rev. B} \href{http://dx.doi.org/10.1103/PhysRevB.37.7306}{{\bf 37} 7306--7325} (1988).

\bibitem{Gumhalter2001}
B.~Gumhalter,
\newblock {\em Physics Reports} \href{http://dx.doi.org/https://doi.org/10.1016/S0370-1573(00)00143-5}{{\bf 351} 1--159} (2001).

\bibitem{Siber2001}
A.~\ifmmode~\check{S}\else \v{S}\fi{}iber, B.~Gumhalter, A.~P. Graham  and J.~P. Toennies,
\newblock {\em Phys. Rev. B} \href{http://dx.doi.org/10.1103/PhysRevB.63.115411}{{\bf 63} 115411} (2001).

\bibitem{Doak1983}
R.~B. Doak, U.~Harten  and J.~P. Toennies,
\newblock {\em Phys. Rev. Lett.} \href{http://dx.doi.org/10.1103/PhysRevLett.51.578}{{\bf 51} 578--581} (1983).

\bibitem{Chis2008}
V.~Chis, B.~Hellsing, G.~Benedek, M.~Bernasconi, E.~V. Chulkov  and J.~P. Toennies,
\newblock {\em Phys. Rev. Lett.} \href{http://dx.doi.org/10.1103/PhysRevLett.101.206102}{{\bf 101} 206102} (2008).

\bibitem{Benedek2010}
G.~Benedek, M.~Bernasconi, V.~Chis, E.~Chulkov, P.~M. Echenique, B.~Hellsing  and J.~{Peter Toennies},
\newblock {\em J. Phys. Condens. Matter} \href{http://dx.doi.org/10.1088/0953-8984/22/8/084020}{{\bf 22} 084020} (2010).

\bibitem{Sklyadneva2011}
I.~Y. Sklyadneva, G.~Benedek, E.~V. Chulkov, P.~M. Echenique, R.~Heid, K.-P. Bohnen  and J.~P. Toennies,
\newblock {\em Phys. Rev. Lett.} \href{http://dx.doi.org/10.1103/PhysRevLett.107.095502}{{\bf 107} 095502} (2011).

\bibitem{Tamtogl2013}
A.~Tamt\"ogl, P.~Kraus, M.~Mayrhofer-Reinhartshuber, D.~Campi, M.~Bernasconi, G.~Benedek  and W.~E. Ernst,
\newblock {\em Phys. Rev. B} \href{http://dx.doi.org/10.1103/PhysRevB.87.035410}{{\bf 87} 035410} (2013).

\bibitem{Benedek2014}
G.~Benedek, M.~Bernasconi, K.~P. Bohnen, D.~Campi, E.~V. Chulkov, P.~M. Echenique, R.~Heid, I.~Y. Sklyadneva  and J.~P. Toennies,
\newblock {\em Phys. Chem. Chem. Phys.} \href{http://dx.doi.org/10.1039/c3cp54834a}{{\bf 16} 7159--7172} (2014).

\bibitem{Manson2016}
J.~R. Manson, G.~Benedek  and S.~Miret-Art{\'{e}}s,
\newblock {\em J. Phys. Chem. Lett.} \href{http://dx.doi.org/10.1021/acs.jpclett.6b00139}{{\bf 7} 1016--1021} (2016).

\bibitem{Benedek2022}
G.~Benedek, J.~R. Manson  and S.~Miret-Art{\'{e}}s,
\newblock {\em Phys. Chem. Chem. Phys.} \href{http://dx.doi.org/10.1039/d2cp03501d}{{\bf 24} 23135--23141} (2022).

\bibitem{Manson2022a}
J.~R. Manson, G.~Benedek  and S.~Miret-Art{\'{e}}s,
\newblock {\em Surf. Sci. Rep.} \href{http://dx.doi.org/10.1016/j.surfrep.2022.100552}{{\bf 77} 100552} (2022).

\bibitem{Schrieffer1957}
J.~Bardeen, L.~N. Cooper  and J.~R. Schrieffer,
\newblock {\em Theory Supercond.} \href{http://dx.doi.org/10.1201/9780429495700}{ 1--332} (1957).

\bibitem{Grimvall1983}
G.~Grimvall,
\newblock {\em Sel. Top. Solid State Phys.} \href{http://dx.doi.org/https://doi.org/10.1002/bbpc.19830870521}{{\bf 16}} (1983).

\bibitem{McMillan1968}
W.~L. McMillan,
\newblock {\em Phys. Rev.} \href{http://dx.doi.org/10.1103/PhysRev.167.326}{{\bf 167} 331--344} (1968).

\bibitem{Allen1975}
P.~B. Allen and R.~C. Dynes,
\newblock {\em Phys. Rev. B} \href{http://dx.doi.org/10.1103/PhysRevB.12.905}{{\bf 12} 905--922} (1975).

\bibitem{Petersen1996}
M.~Petersen, S.~Wilke, P.~Ruggerone, B.~Kohler  and M.~Scheffler,
\newblock {\em Phys. Rev. Lett.} \href{http://dx.doi.org/10.1103/PhysRevLett.76.995}{{\bf 76} 995--998} (1996).

\bibitem{Minniti2012}
M.~Minniti, C.~Díaz, J.~L.~F. Cuñado, A.~Politano, D.~Maccariello, F.~Martín, D.~Farías  and R.~Miranda,
\newblock {\em Journal of Physics: Condensed Matter} \href{http://dx.doi.org/10.1088/0953-8984/24/35/354002}{{\bf 24} 354002} (2012).

\bibitem{Benedek2021}
G.~Benedek, M.~Bernasconi, D.~Campi, I.~V. Silkin, I.~P. Chernov, V.~M. Silkin, E.~V. Chulkov, P.~M. Echenique, J.~P. Toennies, G.~Anemone, A.~A. Taleb, R.~Miranda  and D.~Farías,
\newblock {\em Scientific Reports} \href{http://dx.doi.org/https://doi.org/10.1038/s41598-021-81018-9}{{\bf 11} 1506} (2021).

\bibitem{Hulpke1992}
E.~Hulpke, M.~H\"uppauff, D.-M. Smilgies, A.~D. Kulkarni  and F.~W. de~Wette,
\newblock {\em Phys. Rev. B} \href{http://dx.doi.org/10.1103/PhysRevB.45.1820}{{\bf 45} 1820--1828} (1992).

\bibitem{McMillan2022}
A.~A. McMillan, C.~J. Thompson, M.~M. Kelley, J.~D. Graham, T.~A. Arias  and S.~J. Sibener,
\newblock {\em J. Chem. Phys.} \href{http://dx.doi.org/10.1063/5.0085653}{{\bf 156}} (2022).

\bibitem{Payne1992}
M.~C. Payne, M.~P. Teter, D.~C. Allan, T.~A. Arias  and J.~D. Joannopoulos,
\newblock {\em Rev. Mod. Phys.} \href{http://dx.doi.org/10.1103/RevModPhys.64.1045}{{\bf 64} 1045--1097} (1992).

\bibitem{Sundararaman2017}
R.~Sundararaman, K.~Letchworth-Weaver, K.~A. Schwarz, D.~Gunceler, Y.~Ozhabes  and T.~A. Arias,
\newblock {\em SoftwareX} \href{http://dx.doi.org/10.1016/j.softx.2017.10.006}{{\bf 6} 278--284} (2017).

\bibitem{Schlipf2015}
M.~Schlipf and F.~Gygi,
\newblock {\em Comput. Phys. Commun.} \href{http://dx.doi.org/10.1016/j.cpc.2015.05.011}{{\bf 196} 36--44} (2015).

\bibitem{Note1}
A brief discussion regarding the choice of xc-functional approximation is given in the Supplemental Material, which includes Refs.~\cite {Giustino2017, Marzari2012, Vega2018}.

\bibitem{Perdew2008}
J.~P. Perdew, A.~Ruzsinszky, G.~I. Csonka, O.~A. Vydrov, G.~E. Scuseria, L.~A. Constantin, X.~Zhou  and K.~Burke,
\newblock {\em Phys. Rev. Lett.} \href{http://dx.doi.org/10.1103/PhysRevLett.100.136406}{{\bf 100} 136406} (2008).

\bibitem{Sundararaman2013}
R.~Sundararaman and T.~A. Arias,
\newblock {\em Phys. Rev. B} \href{http://dx.doi.org/10.1103/PhysRevB.87.165122}{{\bf 87} 165122} (2013).

\bibitem{Veit2019}
R.~D. Veit, N.~A. Kautz, R.~G. Farber  and S.~J. Sibener,
\newblock {\em Surf. Sci.} \href{http://dx.doi.org/10.1016/j.susc.2019.06.004}{{\bf 688} 63--68} (2019).

\bibitem{Marzari1997}
N.~Marzari and D.~Vanderbilt,
\newblock {\em Phys. Rev. B} \href{http://dx.doi.org/10.1103/PhysRevB.56.12847}{{\bf 56} 12847--12865} (1997).

\bibitem{Giustino2007}
F.~Giustino, M.~L. Cohen  and S.~G. Louie,
\newblock {\em Phys. Rev. B} \href{http://dx.doi.org/10.1103/PhysRevB.76.165108}{{\bf 76} 165108} (2007).

\bibitem{Brown2016}
A.~M. Brown, R.~Sundararaman, P.~Narang, W.~A. Goddard  and H.~A. Atwater,
\newblock {\em ACS Nano} \href{http://dx.doi.org/10.1021/acsnano.5b06199}{{\bf 10} 957--966} (2016).

\bibitem{Brown2016a}
A.~M. Brown, R.~Sundararaman, P.~Narang, W.~A. Goddard  and H.~A. Atwater,
\newblock {\em Phys. Rev. B} \href{http://dx.doi.org/10.1103/PhysRevB.94.075120}{{\bf 94} 075120} (2016).

\bibitem{Xu2021}
J.~Xu, H.~Takenaka, A.~Habib, R.~Sundararaman  and Y.~Ping,
\newblock {\em Nano Lett.} \href{http://dx.doi.org/10.1021/acs.nanolett.1c03345}{{\bf 21} 9594--9600} (2021).

\bibitem{Xu2021a}
J.~Xu, A.~Habib, R.~Sundararaman  and Y.~Ping,
\newblock {\em Phys. Rev. B} \href{http://dx.doi.org/10.1103/PhysRevB.104.184418}{{\bf 104} 184418} (2021).

\bibitem{hulpke1992helium}
E.~Hulpke and G.~Benedek,
\newblock {\em {Helium Atom Scattering from Surfaces}},
\newblock NATO Asi Series Springer-Verlag (1992).

\bibitem{Meyer1982}
H.-D. Meyer,
\newblock {\em Surface Science} \href{http://dx.doi.org/https://doi.org/10.1016/0039-6028(81)90127-8}{{\bf 104} 117--160} (1981).

\bibitem{Schinke1983}
R.~Schinke and A.~Luntz,
\newblock {\em Surface Science Letters} \href{http://dx.doi.org/https://doi.org/10.1016/0167-2584(83)90773-9}{{\bf 124} L60--L66} (1983).

\bibitem{Celli1984}
V.~Celli, G.~Benedek, U.~Harten, J.~Toennies, R.~Doak  and V.~Bortolani,
\newblock {\em Surface Science} \href{http://dx.doi.org/https://doi.org/10.1016/0039-6028(84)90402-3}{{\bf 143} L376--L382} (1984).

\bibitem{Giustino2017}
F.~Giustino,
\newblock {\em Rev. Mod. Phys.} \href{http://dx.doi.org/10.1103/RevModPhys.89.015003}{{\bf 89} 1--68} (2017).

\bibitem{Marzari2012}
N.~Marzari, A.~A. Mostofi, J.~R. Yates, I.~Souza  and D.~Vanderbilt,
\newblock {\em Rev. Mod. Phys.} \href{http://dx.doi.org/10.1103/RevModPhys.84.1419}{{\bf 84} 1419--1475} (2012).

\bibitem{Vega2018}
L.~Vega, J.~Ruvireta, F.~Viñes  and F.~Illas,
\newblock {\em J. Chem. Theory Comput.} \href{http://dx.doi.org/10.1021/acs.jctc.7b01047}{{\bf 14} 395--403} (2018).

\end{thebibliography}
\end{document}